\documentclass[default,iicol,sn-mathphys]{sn-jnl}% Default with double column layout
\usepackage{subfigure}

\usepackage{geometry}
\usepackage{float}
 \usepackage{etoolbox}
\makeatletter
\patchcmd{\ps@headings}
{\hbox to \hsize{\hfill Springer Nature 2021 \LaTeX\ template\hfill}}
{\hbox to \hsize{}}
{}
{}
\patchcmd{\ps@titlepage}
{\hbox to \hsize{\hfill Springer Nature 2021 \LaTeX\ template\hfill}}
{\hbox to \hsize{}}
{}
{}
\makeatother
\makeatletter
\patchcmd{\ps@headings}
{\hbox to \hsize{\hfill Springer Nature 2021 \LaTeX\ template\hfill}}
{\hbox to \hsize{}}
{}
{}
\patchcmd{\ps@headings}
{\hbox to \hsize{\hfill Springer Nature 2021 \LaTeX\ template\hfill}}
{\hbox to \hsize{}}
{}
{}
\patchcmd{\ps@titlepage}
{\hbox to \hsize{\hfill Springer Nature 2021 \LaTeX\ template\hfill}}
{\hbox to \hsize{}}
{}
{}
\makeatother

\jyear{2021}%

\theoremstyle{thmstyleone}%
\theoremstyle{thmstyletwo}%

\theoremstyle{thmstylethree}%
\usepackage{arydshln}
\usepackage{booktabs}
\usepackage{color}
\usepackage{array}
\usepackage{tabularx}
\usepackage{adjustbox}
\usepackage{mathrsfs}
\usepackage{mathtools}
\usepackage{extarrows}
\usepackage{multirow}
\usepackage{graphicx}
\usepackage{amssymb}
\usepackage{bbding}
\usepackage{overpic}
\usepackage{soul}
\usepackage[marginal]{footmisc}
\usepackage[misc]{ifsym} 

\begin{document}

% \title[Article Title]{Infrared Image Super-Resolution via Lightweight Information Split Network}
\title[ ]{Infrared Image Super-Resolution via Lightweight Information Split Network}
%%                 \dgr{MSc, PhD}}\email{iauthor@gmail.com}Shijie Liu Jie Huang
%%=============================================================%%
\author{ \fnm{Shijie} \sur{Liu}$^{1}$, \fnm{Kang} \sur{Yan}$^{1}$, \fnm{Feiwei} \sur{Qin}$^{1}$, \fnm{Changmiao} \sur{Wang}$^{2}$, \fnm{Ruiquan} \sur{ Ge}$^{1,*}$, \\
\fnm{Kai} \sur{Zhang}$^{3,*}$, \fnm{Jie} \sur{Huang}$^{4}$, \fnm{Yong} \sur{Peng}$^{1}$, \fnm{Jin} \sur{Cao}$^{5}$}

% \author{\fnm{Shijie} \sur{Liu}$^{1}$}
% \author{\fnm{Kang} \sur{Yan}$^{1}$}
% \author{\fnm{Feiwei} \sur{Qin}$^{1}$}
% \author{\fnm{Changmiao} \sur{Wang}$^{2}$}
% \author{\fnm{Ruiquan} \sur{ Ge}$^{1,*}$} 
% \author*{\fnm{Kai} \sur{Zhang}$^{3,*}$}
% \author{\fnm{Jie} \sur{Huang}$^{4}$}
% \author{\fnm{Yong} \sur{Peng}$^{1}$}
% \author{\fnm{Jin} \sur{Cao}$^{5}$}
% \address[author1]{School of Computer Science and Technology, Hangzhou Dianzi University, China}

% \address[author4]{Shenzhen Research Institute of Big Data, China}

% \address[author5]{CVL, ETH Zurich, Switzerland}

\affil[1]{ School of Computer Science and Technology, Hangzhou Dianzi University, China \\
gespring@hdu.edu.cn} 

\affil[2]{Shenzhen Research Institute of Big Data, China}

\affil[3]{\centering CVL, ETH Zurich, Switzerland \\
kai.zhang@vision.ee.ethz.ch}

\affil[4]{School of Information and Electronic Engineering, Zhejiang University of Science and Technology, China}
\affil[5]{Whiting School of Engineering, Johns Hopkins University, America}


\abstract{Single image super-resolution (SR) is an established pixel-level vision task aimed at reconstructing a high-resolution image from its degraded low-resolution counterpart. Despite the notable advancements achieved by leveraging deep neural networks for SR, most existing deep learning architectures feature an extensive number of layers, leading to high computational complexity and substantial memory demands. To mitigate these challenges, we introduce a novel, efficient, and precise single infrared image SR model, termed the Lightweight Information Split Network (LISN). The LISN comprises four main components: shallow feature extraction, deep feature extraction, dense feature fusion, and high-resolution infrared image reconstruction. A key innovation within this model is the introduction of the Lightweight Information Split Block (LISB) for deep feature extraction. The LISB employs a sequential process to extract hierarchical features, which are then aggregated based on the relevance of the features under consideration. By integrating channel splitting and shift operations, the LISB successfully strikes an optimal balance between enhanced SR performance and a lightweight framework. Comprehensive experimental evaluations reveal that the proposed LISN achieves superior performance over contemporary state-of-the-art methods in terms of both SR quality and model complexity, affirming its efficacy for practical deployment in resource-constrained infrared imaging applications. The code is available at \href{https://github.com/sad192/LISN-Infrared-Image-SR}{https://github.com/sad192/LISN-Infrared-Image-SR}}

%%================================%%
%% Sample for structured abstract %%
%%================================%%

% \abstract{\textbf{Purpose:} 
% 
% \textbf{Conclusion:} The abstract serves both as a general introduction to the topic and as a brief, non-technical summary of the main results and their implications. The abstract must not include subheadings (unless expressly permitted in the journal's Instructions to Authors), equations or citations. As a guide the abstract should not exceed 200 words. Most journals do not set a hard limit however authors are advised to check the author instructions for the journal they are submitting to.}


\maketitle
% \renewcommand{\thefootnote}{}
% \footnote[1]{\textrm{\Letter} \fnm{Ruiquan} \sur{ Ge}}

% \footnotetext{gespring@hdu.edu.cn}

% % \renewcommand{\thefootnote}{\fnsymbol{footnote}}
% \footnote[1]{\textrm{\Letter} \fnm{Kai} \sur{Zhang}}
% \renewcommand{\thefootnote}{}
% \footnotetext{kai.zhang@vision.ee.ethz.ch}

% \renewcommand{\thefootnote}{}
% \footnote[1]{$^{1}$\orgaddress{\street{School of Computer Science and Technology, Hangzhou Dianzi University}, \country{China}}}
% \footnote[2]{$^{2}$\orgaddress{\street{Shenzhen Research Institute of Big Data},  \country{China}}}
% \footnote[3]{$^{3}$\orgaddress{\street{CVL, ETH Zurich},  \country{Switzerland}}}
% \footnote[4]{$^{4}$\orgaddress{\street{School of Information and Electronic Engineering, Zhejiang University of Science and Technology},  \country{China}}}
% \footnote[5]{$^{5}$\orgaddress{\street{Whiting School of Engineering, Johns Hopkins University},  \country{America}}}

% \clearpage
\section{Introduction}\label{sec1}
Infrared image super-resolution, a pivotal and persistent area of research, is highly impactful across a broad spectrum of low-level and high-level vision tasks \citep{ming2013effect}. This prominence is attributed to the unique advantages infrared imaging systems present, such as their consistent performance across varying weather conditions and throughout different times of the day. Enhancing the resolution of infrared images allows for the extraction of finer feature information, which, in turn, augments the efficacy of ensuing vision tasks. Visible images, captured within the visible wavelength spectrum, illustrate an object's color and finer details. In contrast, infrared images are monochromatic, focusing solely on thermal radiation intensities emitted by subjects, without revealing color or shading nuances \citep{si2023tri}.
We conducted a detailed analysis of infrared image characteristics, which typically encompass background information devoid of high-frequency semantic features and contain considerable redundancy. To this end, we have implemented a channel splitting operation. However, this operation alone risks impeding cross-channel feature interactions. To strike a balance between performance and parameter quantity, we integrate contrast-aware channel attention \citep{hui2019lightweight}, thereby enhancing the model's representational capacity. We also incorporate a pixel attention block into the feed-forward neural network segment of the conventional Transformer structure to bolster local image structure learning.
In conclusion, we introduce the Lightweight Information Split Network, a novel and efficacious lightweight infrared image SR model that synergistically combines channel splitting operations, CNNs, and shift technologies.
The main contributions of this work can be summarized as follows:
\begin{itemize}
    \item We present a simplified and computationally efficient module, termed the shift block. This innovation is adept at capturing spatial relationships within images while concurrently minimizing computational load.
    \item Through a thorough analysis of infrared image characteristics, we have implemented a channel splitting operation that serves to reduce superfluous computational demands, thus enabling the streamlining of the SR model to achieve a lightweight profile.
    \item We introduce the innovative Lightweight Information Split Block module, which synergistically integrates the shift building block with convolutional neural network techniques to enhance the efficacy of the super-resolution process.
\end{itemize}

\section{Related Work}\label{sec2}

\subsection{Single Visible Image SR}\label{subsec1}
\textbf{Conventional CNNs for SR.} Single-image SR reconstruction aims to retrieve additional detail from a low-resolution or degraded image to enhance its texture clarity. This technique is of substantial significance across various applications in the field of image processing. With the advent of deep learning, a plethora of CNN-based models have been developed, demonstrating notable performance enhancements over traditional image reconstruction methodologies. For instance, He et al. \citep{he2016deep} augmented image SR performance by deepening the neural network architecture. Likewise, Lim et al. \citep{lim2017enhanced} utilized a series of densely interconnected residual blocks to bolster the model's expressive capabilities. Furthermore, Zhang et al. \citep{zhang2020residual} proposed an innovative residual dense network to effectively exploit the hierarchical features intrinsic to the original low-resolution images. Despite the advancements achieved with these CNN-based SR models, they are inherently challenged in capturing comprehensive global contextual information, a limitation stemming from the finite scope of the convolutional kernels. This aspect represents a critical area for potential improvement in the pursuit of more accurate and globally coherent super-resolution reconstruction.

\textbf{Efficient SR.} While numerous prior SR models have exhibited impressive performance, they are often accompanied by considerable computational and memory demands, which limit their practical application. In response to this challenge, various studies have proposed more efficient models for image SR.

Hui et al.~\citep{hui2019lightweight} introduced the Information Multi-Distillation Network (IMDN), a compact architecture that reduces computational burden through the use of multiple channel splitting operations. These operations effectively concentrate informative features, allowing subsequent convolutions to be more computationally efficient. Zhao et al. \citep{zhao2020efficient} developed the Pixel Attention Network (PAN), an SR model notable for its succinctness and efficacy. The model's cornerstone is a novel pixel attention mechanism which, by focusing on relevant pixels, manages to achieve substantial parameter reduction while maintaining high-quality SR results.

Striving to enhance model performance through parameter reduction, HNCT \citep{fang2022hybrid} amalgamates CNN and Transformer architectures. This hybrid approach exploits both local and non-local priors, allowing for the extraction of more robust features essential for SR. Furthermore, BSRN \citep{li2022blueprint} boosts model performance by incorporating blueprint separation convolution, a key technique in crafting efficient CNNs. This method disentangles complex convolution operations into simpler, more manageable components, yielding efficiency gains without compromising effectiveness. Despite the lean and expedient nature of these models, there remains a trade-off: the high-resolution images they produce are often not on par with the quality achieved by their more resource-intensive counterparts. This gap highlights the ongoing need for optimization strategies that can reconcile the dual aims of computational efficiency and superior SR.

%%% 
\textbf{Transformer-based SR.} In recent years, the Transformer, a potent neural network architecture, has attracted considerable attention for its potential in natural language processing. Its exceptional performance across diverse language tasks has prompted researchers to broaden its application, moving beyond language-centric tasks to include those related to vision.

One seminal work by Dosovitskiy et al. \citep{dosovitskiy2020image} leveraged the Vision Transformer in image recognition tasks, successfully addressing classification and localization challenges across extensive image datasets. To adapt the Transformer to handle larger input image sizes, Liang et al. \citep{liang2021swinir} introduced a residual Swin Transformer block, which integrates a convolutional layer at the beginning of the block to amplify feature representation and employs residual connections to facilitate feature aggregation. Chen et al. \citep{chen2021pre} developed a pre-trained Image Processing Transformer that capitalizes on the ViT architecture, employing non-overlapping patch embeddings directly. While Transformer-based methods for SR often surpass those based on CNNs, it is important to recognize the formidable feature extraction capabilities inherent in CNNs.

In this study, we seek to harness the strengths of both Transformer and CNN architectures to formulate an advanced super-resolution algorithm for infrared images. By emphasizing the extraction and integration of both global and local features, our objective is to achieve a more nuanced and effective approach to SR, one that fully exploits the synergistic potential of these two powerful technologies.

\subsection{Single Infrared Image SR}\label{subsec2.1}
Infrared imaging holds significant applications across various fields, and the quest to enhance the quality of infrared images is paramount. Before the advent of advanced machine learning techniques, infrared image enhancement predominantly utilized traditional digital signal processing methods, such as interpolation and wavelet transform. However, the advent of deep learning has shifted the focus towards algorithms that leverage the superior feature extraction capabilities of convolutional neural networks for this purpose.

He et al. \citep{he2018cascaded} introduced a novel SR technique using a cascaded deep network architecture with multiple receptive fields designed to improve the super-resolution process of infrared images, thereby enhancing their quality. Due to the intrinsic differences in imaging systems, the contours and edges in infrared images often appear more blurred when compared to their visible light counterparts. Addressing this challenge, Zou et al. \citep{zou2021super} developed a U-Net-based model employing residual networks to separately extract high and low-frequency information from infrared images, enhancing the detail and clarity of the resulting images.

GAN-based methods have been increasingly prevalent in the super-resolution of visible light images, and this approach has been extended to the realm of infrared imagery. Huang et al. \citep{huang2021infrared} proposed PSRGAN, an efficient model that integrates transfer learning and knowledge distillation to achieve high-quality infrared image super-resolution with a compact model size. Concurrently, Huang et al. \citep{57huang2021infrared} also developed HetSRWGAN which improved the stability of the GAN training process by incorporating a new loss function. Wu et al. \citep{wu2022meta} introduced a lightweight SR method tailored to infrared imaging. This method employs Meta Transfer Learning to facilitate high-resolution reconstruction of infrared images.

\begin{figure*}[t]%
\centering
\begin{overpic}[width=0.9\textwidth]{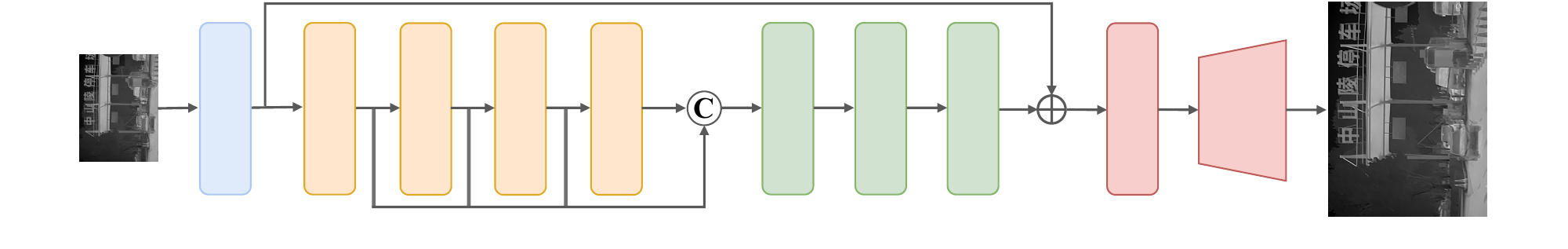}
\put(12.5,0){\color{black}{\footnotesize SFE}}
\put(29.5,0){\color{black}{\footnotesize DFE}}
\put(54.75,0){\color{black}{\footnotesize DFF}}
\put(75,0){\color{black}{\footnotesize IIR}}

\rotatebox{90}{\put(5.75,-14.75){\color{black}{\footnotesize \textbf{Conv3} }}}
\rotatebox{90}{\put(6.75,-6){\color{black}{\footnotesize \textbf{LISB} }}}
\rotatebox{90}{\put(6.75,-5.30){\color{black}{\footnotesize \textbf{LISB} }}}
\rotatebox{90}{\put(6.75,-5.25){\color{black}{\footnotesize \textbf{LISB} }}}
\rotatebox{90}{\put(6.75,-5.4){\color{black}{\footnotesize \textbf{LISB} }}}

\rotatebox{90}{\put(5.75,-10.15){\color{black}{\footnotesize \textbf{Conv1} }}}
\rotatebox{90}{\put(5.75,-5.0){\color{black}{\footnotesize \textbf{Conv3} }}}
\rotatebox{90}{\put(7.5,-5.15){\color{black}{\footnotesize \textbf{PA} }}}

\rotatebox{90}{\put(5.75,-9.40){\color{black}{\footnotesize \textbf{Conv3} }}}
\rotatebox{90}{\put(6.5,-5.25){\color{black}{\footnotesize \textbf{Pixel} }}}
\rotatebox{90}{\put(6.0,-0.25){\color{black}{\footnotesize \textbf{shuffle} }}}

\end{overpic}
\caption{The architecture of the LISN for lightweight infrared image SR. LISN consists of four key components: Shallow Feature Extraction (SFE), Deep Feature Extraction (DFE), Deep Feature Fusion (DFF), and Infrared Image Reconstruction (IIR). 
% $\copyright$ and $\oplus$ represent the concatenate and sum operations, respectively%
}\label{fig1:LISN}
\end{figure*}

\begin{figure*}[t]%
\centering
\begin{overpic}[width=0.9\textwidth]{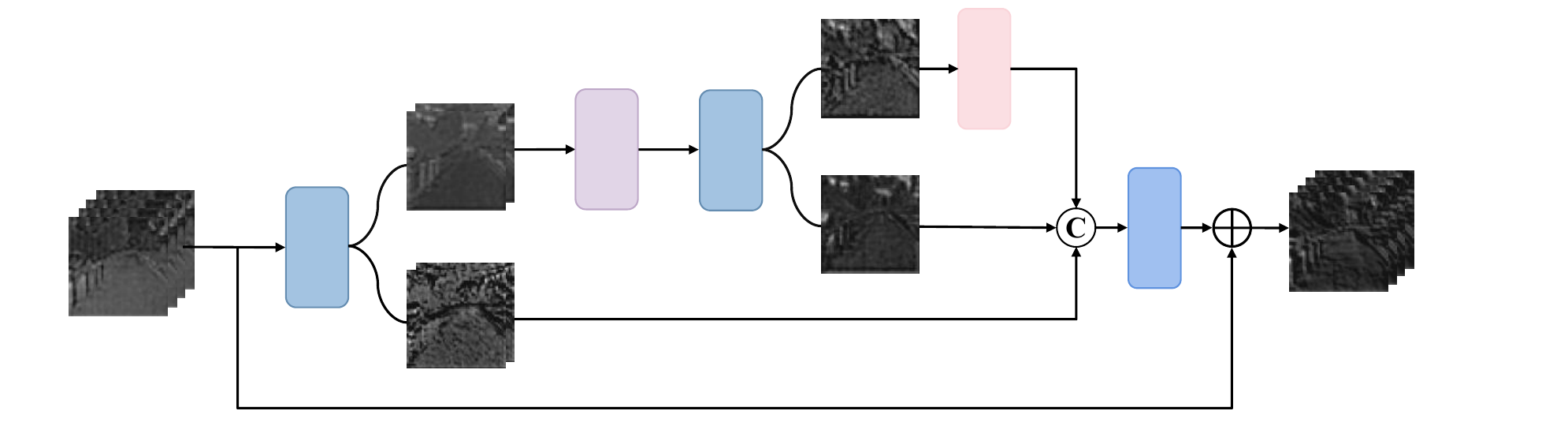}
\put(3.45,5.5){\color{black}{\tiny $H\times W\times C$}}
\put(24.75,2.15){\color{black}{\tiny $H\times W\times C/2$}}
\put(24.75,12.25){\color{black}{\tiny $H \times W \times C/2$}}
\put(51.0,8.25){\color{black}{\tiny $H \times W \times C/4$}}
\put(51.0,18.25){\color{black}{\tiny $H \times W \times C/4$}}
\put(81.25,7.15){\color{black}{\tiny $H \times W \times C$}}

%%%rotate : (y,x)
\rotatebox{90}{\put(8.85,-20.70){\color{black}{\footnotesize \textbf{Split} }}}
\rotatebox{90}{\put(15.40,-17.5){\color{black}{\footnotesize \textbf{SBB} }}}
\rotatebox{90}{\put(15.0,-7.0){\color{black}{\footnotesize \textbf{Split} }}}

\rotatebox{90}{\put(20.15,-15.25){\color{black}{\footnotesize \textbf{RDB} }}}
\rotatebox{90}{\put(10.15,-10.00){\color{black}{\footnotesize \textbf{CCA} }}}

\end{overpic}
\caption{The architecture of the LISB for deep feature extraction. LISB is primarily composed of the following modules: Shift Building Block (SBB), Residual Depth-wise Convolution Block (RDB), Contrast-aware Channel Attention Block (CCA)}\label{fig2:LISB}
\end{figure*}
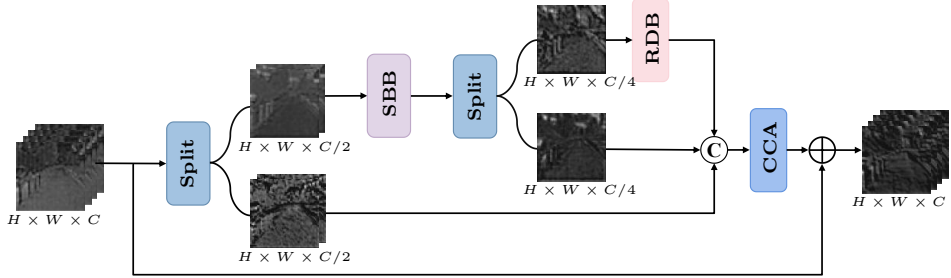

\section{Method}\label{sec3}
\subsection{ Network Architecture}\label{subsec2}
In light of recent advancements and the increasing demand for compact yet high-performing models, we present a novel network architecture in this section. As illustrated in Fig. \ref{fig1:LISN}, our proposed LISN consists of four key components: shallow feature extraction, deep feature extraction, dense feature fusion, and high-resolution (HR) infrared image reconstruction modules. The network initiates the processing pipeline with a convolutional layer, adept at performing early visual feature extraction. This layer not only facilitates smoother optimization but also contributes to enhanced performance outcomes. It efficiently maps the input image from its original space into a higher-dimensional feature space, allowing for more intricate and expressive feature representations. Given a low-resolution infrared image, denoted as $\mathcal{I_{LR}} \in \mathbb{R}^{H \times W \times C_{in}}$, where $\mathcal{H}$, $\mathcal{W}$, and $C_{in}$ signify the image's height, width, and number of input channels respectively, we employ a $3 \times 3$ convolutional layer to carry out the initial shallow feature extraction. 
\begin{equation}
\mathcal{F}_0=\mathcal{H_{S F}}\left(I_{L R}\right)=k^{3 \times 3} \left( I_{L R} \right),
\end{equation}
where $\mathcal{F}_0 \in \mathbb{R}^{H \times W \times C_{out}}$ represents the shallow features extracted by the initial $3 \times 3$ convolutional layer, denoted as $\mathcal{H_{SF}}$. Here, $C_{out}$ indicates the number of output channels, and $k^{3 \times 3}$ refers to the convolutional kernel size of $3 \times 3$. The extracted shallow feature, $\mathcal{F}_0$, serves as the input to a deep feature extraction module composed of multiple Lightweight Information Split Blocks. Given that the total number of LISBs in the module is denoted by $N$, the output feature map of the $n$-th LISB, $\mathcal{F}_{n}$ (where $1 \leq n \leq N$), is formulated as follows:
\begin{equation}
\mathcal{F}_{n}=\left\{\begin{array}{ll}
\mathcal{LB}_{n}\left(\mathcal{F}_{n-1}\right) & n\ge1, \\
\mathcal{F}_0 & n=0,
\end{array}\right.
\label{deep-2}
\end{equation}
where $\mathcal{LB}_{n}$ denotes the function of n-$\mathrm{th}$ LISB. More details about the LISB are described further in Section \ref{LISB}.

The outputs of all the LISBs are concatenated and relayed to the dense feature fusion module. This critical module amalgamates the entire hierarchical feature set by employing a sequential arrangement of a $1 \times 1$ convolutional layer and a $3 \times 3$ convolutional layer. It also incorporates a pixel attention (PA) mechanism, as introduced by Zhao et al. \citep{zhao2020efficient}, which computes attention coefficients for every pixel within the feature map. The PA mechanism operates analogously to the concepts of channel attention and spatial attention in terms of its functional approach. Distinctively, PA generates three-dimensional attentional maps instead of conventional one-dimensional attentional vectors or two-dimensional attention maps. This attention scheme is particularly noteworthy for its minimal introduction of additional parameters while concurrently achieving enhanced SR results for infrared images, leading to an output that is mathematically formulated as follows:
\begin{equation}
\mathcal{F}_{out}= M_{p}\left( k^{3 \times 3} \left(k^{1 \times 1} \left[\mathcal{F}_{1}, \mathcal{F}_{2}, \cdots, \mathcal{F}_{N}\right]\right)\right),
\end{equation}
where $\mathcal{F}_{out}$ is the final output of the dense feature fusion module. The operation $M_{p}$ denotes the pixel attention layer that modulates features spatially, applying attention to each pixel individually. The notation $[\cdot, \cdot]$ represents the concatenation operation, which merges feature maps along the channel dimension. In the concluding phase of the network's processing, we achieve the reconstruction of the HR infrared image, $I_{SR}$, by skillfully merging both shallow and deep features. The reconstruction process is formalized as follows:
\begin{equation}
I_{S R}=\mathcal{U}\left(\mathcal{F}_{out} + \mathcal{F}_{0}\right),
\end{equation}
where $\mathcal{U}(\cdot)$ signifies the module responsible for HR infrared images. This module is realized through the deployment of a $3 \times 3$ convolutional layer followed by a pixel shuffle layer, which effectively up-samples the feature maps. By integrating the shallow feature $\mathcal{F}_{0}$, the reconstruction module is designed to produce an HR infrared image endowed with enhanced sharpness and improved texture detail.

To further enhance the structural fidelity of the reconstruction target and improve the model's SR reconstruction capabilities, we incorporate the Sobel operator. This operator is employed to compute the edge loss between the HR image and the SR image, thus emphasizing the preservation of edge information which is crucial for perceptual quality. The loss function for our LISN is articulated as follows:
\begin{equation}
\hspace{-2mm}
\mathcal{L}=\left\|I_{S R}-I_{H R}\right\|^{1} + \alpha_{1} \ast \left\|\mathcal{S}\left(I_{SR}\right) -\mathcal{S}\left(I_{HR}\right)\right\|^{1},
\end{equation}
where $\alpha_{1}$ is a weighting parameter that adjusts the contribution of the edge loss to the overall loss function; empirically, we have determined its optimal value to be 0.1. The operator $\left\|\right\|^{1}$ denotes the $L_{1}$ norm, which serves as a measure of reconstruction error. Within this context, $I_{SR}$ represents the super-resolved image obtained when the low-resolution input $I_{LR}$ is processed through the LISN, while $I_{HR}$ corresponds to the ground-truth high-resolution image. Furthermore, $\mathcal{S}(\cdot)$ symbolizes the Sobel operator, a differential operator used for edge detection.

% module img 
\begin{figure*}[htbp]%
\centering
\begin{overpic}[width=0.9\textwidth]{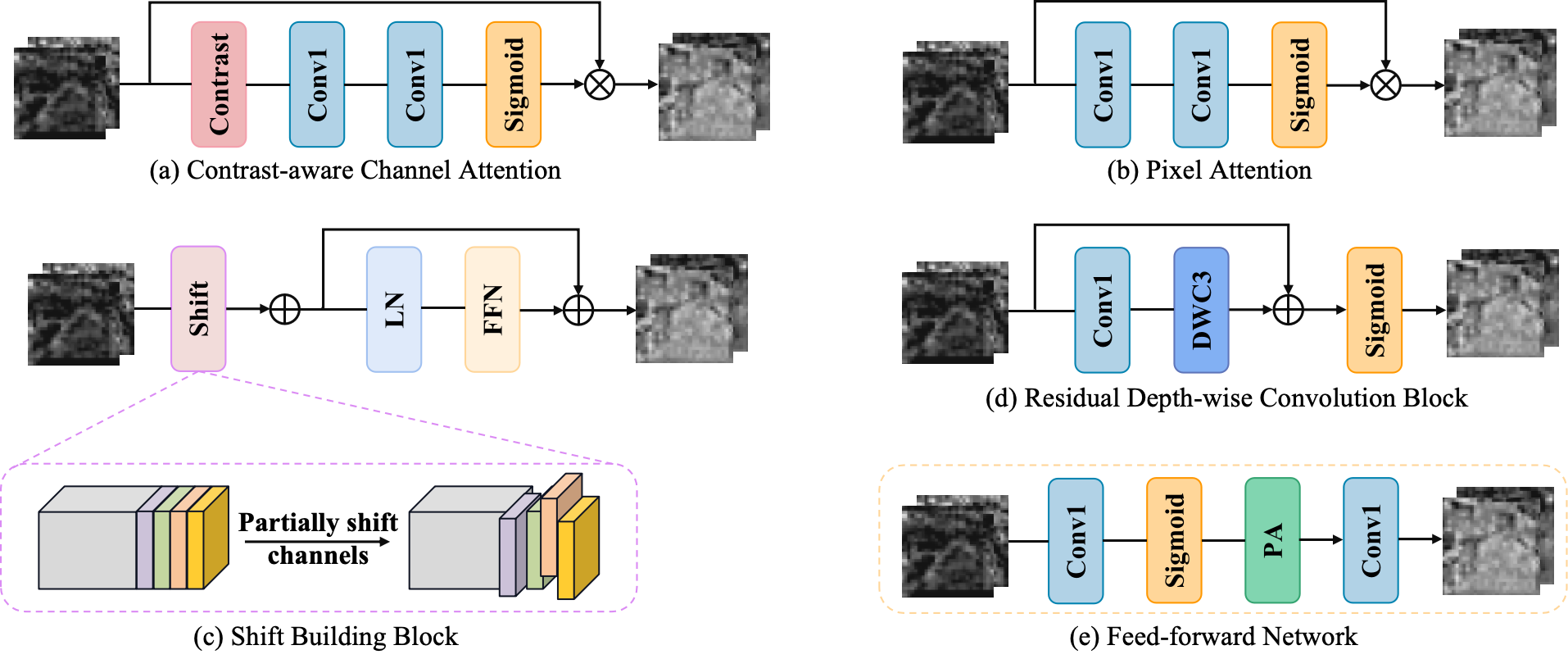}

\end{overpic}
\caption{This demonstrates the main components of LISB. (a) The architecture of CCA. (b) The architecture of PA. (c) The architecture of shift block. (d) The architecture of RDB. The DWC3 means that the kernel size is $3 \times 3$ for depth-wise convolution.}\label{fig3:part of LISN}
\end{figure*}

\subsection{Lightweight Information Split Block} \label{LISB}
As depicted in Fig. \ref{fig2:LISB}, the LISB is composed of several key elements: the channel splitting operation, a single Shift Building Block (SBB), a Residual Depth-wise Convolution Block (RDB), and a Contrast-aware Channel Attention (CCA) module. The SBB consists of two shift Transformer layers, which employ zero-parameter and zero-FLOP shift operations to efficiently capture spatial relationships, significantly improving computational efficiency. The shift operation moves input feature maps along the spatial axes, left, right, up, and down, by shifting a small fraction of the channels. This deliberate displacement facilitates the direct mixing of information among adjacent features, enhancing feature representation. To underscore the importance of local contextual information, the RDB module is incorporated, which includes a $1\times1$ convolution, a $3\times3$ depth-wise convolution, and is followed by a Sigmoid activation function. The CCA module serves to model inter-channel correlations, autonomously learning the significance of each channel. As a result, this process dynamically weights each channel, amplifying important features and attenuating less critical ones.

According to Eq.~\ref{deep-2}, the feature maps of the (n-1)-th LISB, denoted by $\mathcal{F}_{n-1}$, are fed directly into the n-th LISB. Given $\mathcal{F}_{n-1}$ as input, the n-th LISB begins by employing the channel splitting layer to bifurcate the input into two halves, as described by:
\begin{equation}
r_{1}, m_{1} = \mathcal{C_{S}}\left(\mathcal{F}_{n-1}\right),
\end{equation}
where $\mathcal{C_{S}}(\cdot)$ symbolizes the channel splitting operation, with $r_{1}$ and $m_{1}$ being the resultant split feature maps.

Following this, the SBB processes $m_{1}$ and further separates the intermediate features into $r_{2}$ and $m_{2}$ using another channel splitting layer. These feature maps, $r_{2}$ and $m_{2}$, are obtained as per the following set of operations:
\begin{equation}
\begin{array}{l}
m_{1} = \operatorname{Shift}(m_{1}), \\
m_{1} = \operatorname{FFN}(\operatorname{LN}(m_{1})) + m_{1}, \\
r_{2}, m_{2} = \mathcal{C_{S}}\left(m_{1}\right),
\end{array}
\end{equation}
wherein 'Shift' and 'FFN' denote the shift operation and feed-forward network, respectively. A LayerNorm (LN) precedes the FFN to normalize the features.

Subsequently, $m_{2}$ is introduced to the Residual Depth-wise Convolution Block, which leverages residual learning to further refine the intermediate features, yielding $m_{3}$. The computation of the feature map $m_{3}$ is as follows:
\begin{equation}
m_{3} = \sigma\left( k^{1 \times 1}\left( k^{3 \times 3}\left( m_{2} \right) \right) + m_{2} \right),
\end{equation}
where $\sigma$ represents the Sigmoid activation function, and $k^{a \times a}$ corresponds to a convolutional operation with a kernel size of $a \times a$.

Upon extracting $m_{3}$, the feature maps $r_{1}$, $r_{2}$, and $m_{3}$ are concatenated and fed into the CCA layer. The purpose of the CCA layer is to amplify the channel features that are pertinent to the task of infrared image reconstruction. Additionally, $\mathcal{F}_{n-1}$ is combined with the output of the CCA layer via a long skip connection. This facilitates the deep feature extraction module in focusing on the processing of high-frequency information and contributes to the stabilization of the training process.

Assuming the input to this stage is denoted as $x_{in} = [r_{1}, r_{2}, m_{3}] = [x_{1}, \dots, x_{c}, \dots, x_{\mathcal{C}}]$, which comprises $\mathcal{C}$ feature maps each of spatial dimensions $h \times w$. The equation for the concluding step in the deep feature extraction sequence is illustrated as follows:
\begin{equation}
% \hspace{-2mm}
\begin{array}{ll}
t_{1}=\sqrt{\frac{1}{hw} \sum_{c=1}^{\mathcal{C} }\sum_{(i, j) \in x_{c}}\left(x_{c}^{i, j}-\frac{1}{hw} \sum_{(i, j) \in x_{c}} x_{c}^{i, j}\right)^{2}}\\
+ \mathcal{P}_{avg}(x_{in}),

\\[2mm]

\mathcal{F}_{n}=\sigma\left(k^{1 \times 1}\left( k^{1 \times 1} \left( t_{1}  \right)\right)\right)\ast x_{in}+\mathcal{F}_{n-1},
\end{array}
\end{equation}

where $\mathcal{P(\cdot)}$ denotes average pooling, $t_{1}$ represents the sum of the output of the CCA module contrast operation and the average pooling layer output.

\subsection{Residual Depth-wise Convolution Block}\label{subsec4}
Infrared image super-resolution is a task that demands precision at the pixel level. To address this requirement, we introduce the RDB. As illustrated in Fig. \ref{fig3:part of LISN} {\color{blue}(d)}, the RDB uniquely combines the principles of residual learning with depth-wise convolution. This design choice not only mitigates the loss of information but also serves to extract an abundance of local details, which is essential for the enhancement of infrared images. Additionally, the incorporation of depth-wise convolution in the RDB architecture significantly reduces the overall parameter count. Following the extraction of global high-dimensional semantic information by the SBB, the RDB is tasked with honing in on local texture details, thereby contributing to the superior reconstruction quality of infrared images.

\subsection{Attention Mechanism}\label{subsec5}
%% 增加对 attention的描述 参照2.2 $h\ast w \ast c$
The attention mechanism plays a crucial role in emphasizing specific features by allocating increased weights to them, thereby granting these features greater precedence during the processing of subsequent data. The Squeeze-and-Excitation Network (SE-Net), as proposed by Hu et al. \citep{hu2018squeeze}, pioneered the dynamic modulation of inter-channel relationships, surpassing the performance of previous convolutional neural network architectures. 

Building upon this concept, the Convolutional Block Attention Module (CBAM) introduced by Woo et al. \citep{woo2018cbam} utilizes a two-dimensional convolutional layer with a kernel size of $k \times k$ to calculate spatial attention. This spatial attention is then integrated with channel attention to form a composite three-dimensional attention map. To balance model efficiency with computational demand, the PAN proposed by Zhao et al. \citep{zhao2020efficient} presents a straightforward yet effective pixel attention module that achieves enhanced results in image super-resolution tasks.

Given the proven efficacy of these attention mechanisms in boosting the performance of deep learning models for computer vision applications, we have incorporated two distinct attention mechanisms, CCA and PA, into our work. As depicted in Fig. \ref{fig3:part of LISN} {\color{blue}(a)}, CCA is employed after initial deep feature extraction to compute a contrastive loss between pairs of feature maps. This loss metric is then used to pinpoint the most influential channels within the deep feature map. Subsequently, the PA module is introduced to further refine the weightings of high-dimensional semantic information, culminating in an enhanced reconstruction of infrared images. The PA module is capable of generating an attention map with dimensions $h \times w \times c$, the architecture of which is presented in Fig. \ref{fig3:part of LISN} {\color{blue}(b)}.

\section{Experiments}\label{sec4}
\subsection{Datasets and Metrics}\label{subsec6}
For our experiments, we employed the CVC-09-1K dataset, which consists of 1,000 infrared images selected from the CVC-09: FIR Sequence Pedestrian Dataset \citep{cvc09socarras2013adapting}. This dataset was randomly partitioned into two subsets for training and testing purposes, adhering to an 80:20 split. To ensure a comprehensive analysis, we conducted evaluations across two distinct datasets, referred to as results-A \citep{Aliu2018infrared} and results-C \citep{Czhang2017infrared}. The imagery for these datasets was synthesized by fusing infrared and visible-spectrum images using methods cited in the existing literature.

The quality of super-resolved infrared images is conventionally measured using two established metrics: Peak Signal-to-Noise Ratio (PSNR) and Structural Similarity (SSIM). Accordingly, we have adopted these metrics to quantify the performance of our SR methods, enabling a robust assessment of the reconstructed infrared images. Among these metrics, PSNR is utilized to quantify the disparity in pixel values between two images, assessing the level of noise or distortion in the reconstructed image by comparing it with the original. A higher PSNR value indicates a closer resemblance between the reconstructed and original images. On the other hand, SSIM is another commonly employed metric that considers structural details, offering a more refined evaluation of perceived image quality. By analyzing local patterns, brightness, and contrast in both the original and reconstructed images, SSIM assesses the fidelity of the resulting image.

\subsection{Implementation Details}
We generate low-resolution training images by down-sampling HR images using bicubic interpolation with scaling factors of $\times2$ and $\times4$. The resulting LR images undergo augmentation through random rotations of $90^{\circ}$, $180^{\circ}$, and $270^{\circ}$, as well as horizontal flips. The edge loss weight is set to 0.1 by default. For optimization, we employ the Adam optimizer with hyperparameters set to $\beta_1 = 0.9$, $\beta_2 = 0.999$, and $\varepsilon = 10^{-8}$. Our proposed LISN is trained over a total of 6000 epochs. The learning rate is initially set at $2 \times 10^{-4}$ and is reduced by half every 200 epochs. After reaching 1000 epochs, the learning rate is reset to its original value. To strike an optimal balance between the number of parameters and model performance, our LISN comprises only six LISBs. The implementation of LISN is carried out using the PyTorch framework on an NVIDIA 2080Ti GPU.
\begin{table*}[ht]
\begin{center}\small
\caption{Quantitative comparison (average PSNR/SSIM) with state-of-the-art methods for scale factor 2 and 4 on three public datasets. }\label{tabResult}
\begin{tabular*}{\textwidth}{@{\extracolsep{\fill}}lcccccccc@{\extracolsep{\fill}}}
\toprule%
 \multirow{2}*{Method} & \multirow{2}*{Scale} & \multirow{2}*{Params (K)} & \multicolumn{2}{@{}c@{}}{CVC-09-1K} &\multicolumn{2}{@{}c@{}}{results-A} &\multicolumn{2}{@{}c@{}}{results-C} \\\cmidrule{4-5} \cmidrule{6-7} \cmidrule{8-9}%
~ & ~ & ~ & PSNR & SSIM & PSNR & SSIM & PSNR & SSIM\\
\midrule
IMDN \citep{hui2019lightweight}  & \multirow{8}{*}{x4} & 715 & 39.52 & 0.9279 & 33.15 & 0.8321 & 33.82 & 0.8503 \\
% IMDN\citep{hui2019lightweight} &  & 715K  & 39.52  & 0.9279  \\
BSRN \citep{li2022blueprint} &  & 352  & 39.37  & 0.9267 & 33.11  & 0.8311 & 33.80  & 0.8502 \\
PAN \citep{zhao2020efficient} &  & 272  & 39.35  & 0.9262 & 33.13 &0.8315 & 33.82 & 0.8506\\
PSRGAN \citep{huang2021infrared} &  & 349  & 38.73  & 0.9176 & 32.67 & 0.8029 & 33.35  & 0.8038 \\
SwinIR \citep{liang2021swinir} &  & 897  & 39.58  & 0.9309 &\textbf{33.23} & \textbf{0.8332} &33.85 &0.8511 \\
HNCT \citep{fang2022hybrid} &  & 372  & 39.50  & 0.9271 & 33.15  & 0.8312 & 33.82  & 0.8501 \\
\textbf{LISN (Ours) }&  & 279 & \textbf{39.70} & \textbf{0.9317} & 33.16 & 0.8316 & \textbf{33.86} & \textbf{0.8513}\\
\midrule
IMDN \citep{hui2019lightweight}  & \multirow{8}{*}{x2} & 694 & 43.43 & 0.9620 & 37.86 & 0.9334 & 38.85 & 0.9464 \\
BSRN \citep{li2022blueprint} &  & 332  & 43.14  & 0.9583 &37.95 &0.9337 &38.83 &0.9459\\
PAN \citep{zhao2020efficient} &  & 261  & 43.10  & 0.9581 &37.92 &0.9333 &38.81 & 0.9456\\
PSRGAN \citep{huang2021infrared} &  & 312  & 42.38  & 0.9304 & 37.39 & 0.8945 & 38.42 & 0.8987\\
SwinIR \citep{liang2021swinir} &  & 878  & 43.57  & 0.9633 & \textbf{37.96}  & \textbf{0.9351} & 38.91  & 0.9470 \\
HNCT \citep{fang2022hybrid} &  & 356  & 43.39  & 0.9622 & 37.87  & 0.9339 & 38.82  & 0.9457  \\
\textbf{LISN (Ours) }&  & 258 & \textbf{43.65}  & \textbf{0.9642} & 37.94 & 0.9347 & \textbf{38.92} & \textbf{0.9475}\\
\bottomrule
\end{tabular*}
\end{center}
\end{table*}

%%%%%%%%%%%% visual images
\begin{figure*}[htbp]
% \footnotesize \resizebox{\textwidth}{60mm}
\centering
\resizebox{\textwidth}{!}{
\begin{tabular}{cccc}
% this is the first column
\hspace{-2mm}
\begin{adjustbox}{valign=t}
    \begin{tabular}{c}
    \specialrule{0em}{0em}{0pt}
    \includegraphics[width=0.35\textwidth]{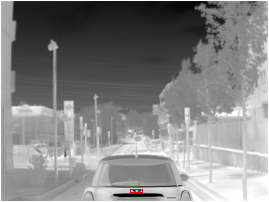}\\ 
    \specialrule{0em}{2pt}{0pt}
    CVC-09-1K: 022822 ($\times$4) \\
    
    \specialrule{0em}{0pt}{7pt}
    \includegraphics[width=0.35\textwidth]{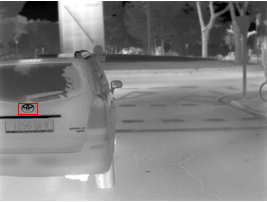} \\
    \specialrule{0em}{2pt}{0pt}
    CVC-09-1K: 000980 ($\times$4)\\
%%%%%%%%%%%%%%%%%%%%%%%%%%%%%%%%%%%%%%%%%%%%%%%%%%%%%%%%%%%%
    \specialrule{0em}{0pt}{10pt}
    \includegraphics[width=0.35\textwidth]{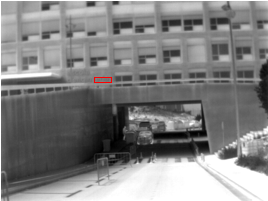} \\
    \specialrule{0em}{2pt}{0pt}
    CVC-09-1K: 027655 ($\times$2)\\
    
    \specialrule{0em}{0pt}{7pt}
    \includegraphics[width=0.35\textwidth]{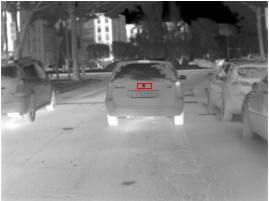} \\
    \specialrule{0em}{2pt}{0pt}
    CVC-09-1K: 001329 ($\times$2)\\
    \end{tabular}
\end{adjustbox}
\hspace{-4mm} 
\begin{adjustbox}{valign=t}
\begin{tabular}{cccc}
\specialrule{0em}{0pt}{0cm}
\includegraphics[width=0.15\textwidth,height=1.5cm]{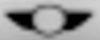}\hspace{-3mm} & 
\includegraphics[width=0.15\textwidth,height=1.5cm]{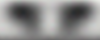}\hspace{-3mm} & 
\includegraphics[width=0.15\textwidth,height=1.5cm]{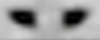}\hspace{-3mm} & \includegraphics[width=0.15\textwidth,height=1.5cm]{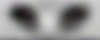}\hspace{-1mm}\\
\specialrule{0em}{-3pt}{0pt}
HR\hspace{-3mm} & IMDN \hspace{-3mm} & BSRN \hspace{-3mm} & PAN \hspace{-3mm}\\
\specialrule{0em}{-3pt}{0pt}
PSNR/SSIM\hspace{-3mm} & 36.29/0.9456 \hspace{-3mm} & 36.69/0.9466 \hspace{-3mm} & 35.54/0.9401 \hspace{-3mm}\\
%%% 
\specialrule{0em}{-2pt}{0pt}
\includegraphics[width=0.15\textwidth,height=1.5cm]{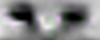}\hspace{-3mm} & 
\includegraphics[width=0.15\textwidth,height=1.5cm]{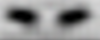}\hspace{-3mm} & \includegraphics[width=0.15\textwidth,height=1.5cm]{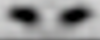}\hspace{-3mm} & \includegraphics[width=0.15\textwidth,height=1.5cm]{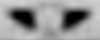}\hspace{0mm} \\
\specialrule{0em}{-3pt}{0pt}
PSRGAN\hspace{-3mm} & SwinIR\hspace{-3mm} & HNCT\hspace{-3mm} & LISN \hspace{-3mm}\\
\specialrule{0em}{-3pt}{0pt}
30.38/0.8841\hspace{-3mm} & 36.31/0.9453\hspace{-3mm} & 36.17/0.9433\hspace{-3mm} & \textbf{37.48/0.9467}\hspace{-3mm}\\

%%%%%%%%%%%%%%%%%%%%%%%%%%%%%%%%%%%%%%%%%%%%%%%%%%%%%%%%%%%%%%%%%%%%%%%%%%%%%%%%%%%%%%%%%%%%%%%%%%%%%%%%%
\specialrule{0em}{0pt}{18pt}
\includegraphics[width=0.15\textwidth,height=1.5cm]{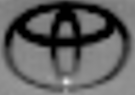}\hspace{-3mm} & 
\includegraphics[width=0.15\textwidth,height=1.5cm]{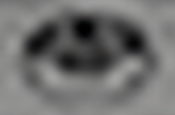}\hspace{-3mm} & 
\includegraphics[width=0.15\textwidth,height=1.5cm]{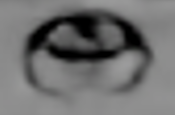}\hspace{-3mm} & 
\includegraphics[width=0.15\textwidth,height=1.5cm]{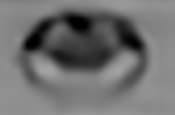}\hspace{-1mm} \\
\specialrule{0em}{-3pt}{0pt}
HR\hspace{-3mm} & IMDN \hspace{-3mm} & BSRN \hspace{-3mm} & PAN \hspace{-3mm} \\
\specialrule{0em}{-3pt}{0pt}
PSNR/SSIM\hspace{-3mm} & 36.69/0.9256 \hspace{-3mm} & 36.77/0.9259 \hspace{-3mm} & 36.34/0.9221 \hspace{-3mm}\\
\specialrule{0em}{-2pt}{0pt}
\includegraphics[width=0.15\textwidth,height=1.5cm]{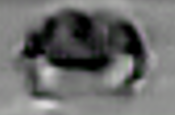}\hspace{-3mm} & 
\includegraphics[width=0.15\textwidth,height=1.5cm]{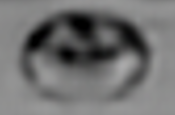}\hspace{-3mm} & \includegraphics[width=0.15\textwidth,height=1.5cm]{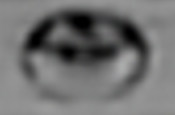}\hspace{-3mm} & \includegraphics[width=0.15\textwidth,height=1.5cm]{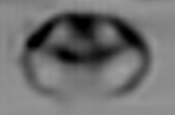}\hspace{0mm} \\
\specialrule{0em}{-3pt}{0pt}
PSRGAN\hspace{-3mm} & SwinIR\hspace{-3mm} & HNCT\hspace{-3mm} & LISN \hspace{-3mm}\\
\specialrule{0em}{-3pt}{0pt}
33.61/0.8703\hspace{-3mm} & 36.76/0.9216\hspace{-3mm} & 36.70/0.9257\hspace{-3mm} & \textbf{36.87/0.9271}\hspace{-3mm}\\
%%%%%%%%% x2
\specialrule{0em}{0pt}{21pt}
\includegraphics[width=0.15\textwidth,height=1.5cm]{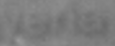}\hspace{-3mm} & 
\includegraphics[width=0.15\textwidth,height=1.5cm]{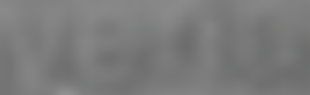}\hspace{-3mm} & 
\includegraphics[width=0.15\textwidth,height=1.5cm]{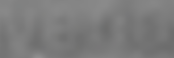}\hspace{-3mm} &
\includegraphics[width=0.15\textwidth,height=1.5cm]{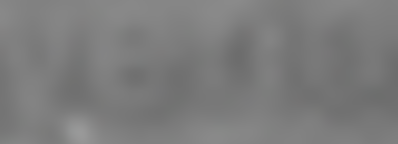}\hspace{-3mm}\\
\specialrule{0em}{-3pt}{0pt}
HR\hspace{-3mm} & IMDN \hspace{-3mm} & BSRN \hspace{-3mm} & PAN \hspace{-3mm} \\
\specialrule{0em}{-3pt}{0pt}
PSNR/SSIM\hspace{-3mm} & 42.97/0.9670 \hspace{-3mm} & 41.77/0.9667 \hspace{-3mm} & 42.19/0.9699 \hspace{-3mm}\\
\specialrule{0em}{-2pt}{0cm}
\includegraphics[width=0.15\textwidth,height=1.5cm]{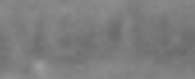}\hspace{-3mm} & 
\includegraphics[width=0.15\textwidth,height=1.5cm]{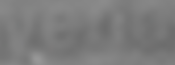}\hspace{-3mm} & \includegraphics[width=0.15\textwidth,height=1.5cm]{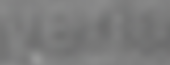}\hspace{-3mm} & \includegraphics[width=0.15\textwidth,height=1.5cm]{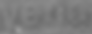}\hspace{-3mm}\\
\specialrule{0em}{-3pt}{0pt}
PSRGAN\hspace{-3mm} & SwinIR\hspace{-3mm} & HNCT\hspace{-3mm} & LISN \hspace{-3mm}\\
\specialrule{0em}{-3pt}{0pt}
38.93/0.9448\hspace{-3mm} & 42.26/0.9700\hspace{-3mm} & 42.21/0.9699\hspace{-3mm} & \textbf{42.32/0.9709}\hspace{-3mm}\\

%%%%
\specialrule{0em}{0pt}{18pt}
\includegraphics[width=0.15\textwidth,height=1.5cm]{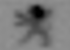}\hspace{-3mm} & 
\includegraphics[width=0.15\textwidth,height=1.5cm]{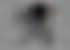}\hspace{-3mm} & 
\includegraphics[width=0.15\textwidth,height=1.5cm]{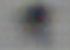}\hspace{-3mm} &
\includegraphics[width=0.15\textwidth,height=1.5cm]{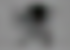}\hspace{-1mm}
\\
\specialrule{0em}{-3pt}{0pt}
HR\hspace{-3mm} & IMDN \hspace{-3mm} & BSRN \hspace{-3mm} & PAN \hspace{-3mm} \\
\specialrule{0em}{-3pt}{0pt}
PSNR/SSIM\hspace{-3mm} & 42.23/0.9697 \hspace{-3mm} & 42.32/0.9671 \hspace{-3mm} & 42.64/0.9658 \hspace{-3mm}\\
\specialrule{0em}{-2pt}{0cm}
\includegraphics[width=0.15\textwidth,height=1.5cm]{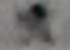}\hspace{-3mm} & 
\includegraphics[width=0.15\textwidth,height=1.5cm]{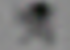}\hspace{-3mm} & 
\includegraphics[width=0.15\textwidth,height=1.5cm]{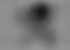}\hspace{-3mm} &  \includegraphics[width=0.15\textwidth,height=1.5cm]{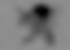}\hspace{0mm}
\\
\specialrule{0em}{-3pt}{0pt}
PSRGAN\hspace{-3mm} & SwinIR\hspace{-3mm} & HNCT\hspace{-3mm} & LISN \hspace{-3mm}\\
\specialrule{0em}{-3pt}{0pt}
39.16/0.9336\hspace{-3mm} & 42.83/0.9663\hspace{-3mm} & 42.84/0.9667\hspace{-3mm} & \textbf{42.91/0.9674}\hspace{-3mm}\\

\end{tabular}
\end{adjustbox}
\end{tabular}}
% \vspace{-2mm}
\caption{
Visualization comparison of different pruning methods on CVC-09-1K dataset. 
}
\label{fig: visual compare1}
\vspace{-4mm}
\end{figure*} 

\subsection{Results on Infrared Image SR}\label{subsec7}
We conducted a comparative evaluation of our proposed LISN with several state-of-the-art SR methods, including IMDN \citep{hui2019lightweight}, BSRN \citep{li2022blueprint}, PAN \citep{zhao2020efficient}, PSRGAN \citep{huang2021infrared}, SwinIR \citep{liang2021swinir}, and HNCT \citep{fang2022hybrid}. Table \ref{tabResult} presents the quantitative results for different upscaling factors, showcasing the comparative analysis.

Our LISN demonstrates exceptional performance with a considerably lower parameter count when contrasted with other SR techniques. On the CVC-09-1K dataset, LISN achieves the highest improvements with gains of 0.97 in PSNR and 0.0141 in SSIM for the $\times4$ upscaling factor, respectively. Similarly, during the implementation of $\times2$ infrared image super-resolution reconstruction, LISN also demonstrates superior reconstruction performance on the CVC-09-1K dataset. Specifically, compared to the comparative methods, LISN achieves the maximum increase of 1.27 in PSNR and 0.0338 in SSIM. While the SR performance of HNCT and SwinIR for infrared images is comparable to that of our LISN method, the number of parameters for these models is approximately 93K and 618K higher, respectively, in all cases.

\begin{figure*}[t]
\centering
\subfigure[PSNR vs. Epochs ($\times4$)]
{
    \begin{minipage}[b]{.45\textwidth}
        \centering
        \begin{overpic}[scale=0.45]{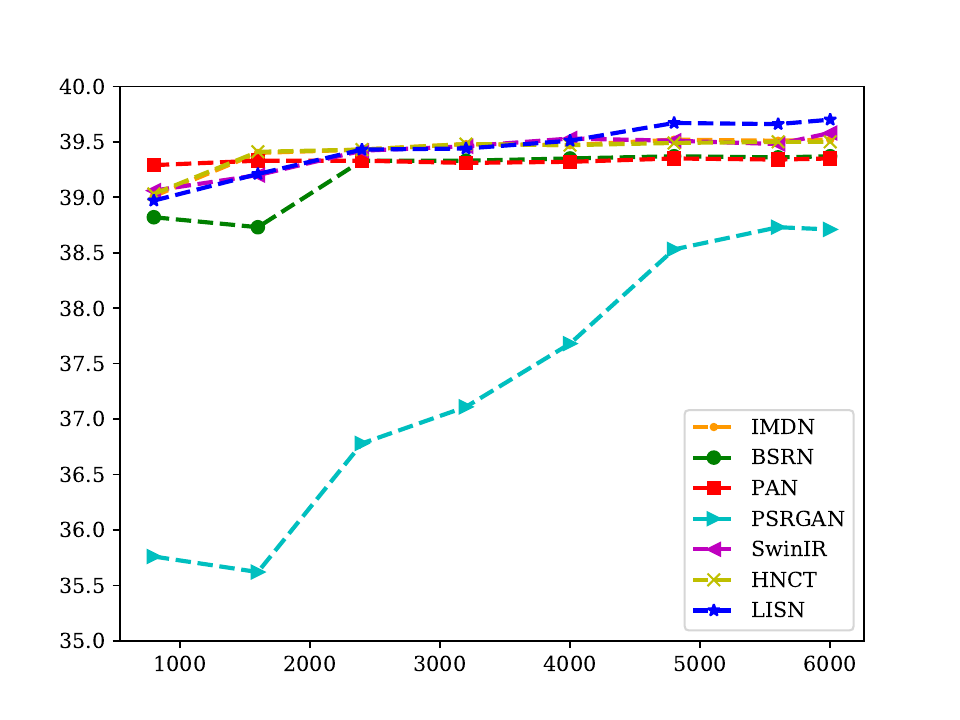}
        \put(25,-0.85){\color{black}{\footnotesize Number of training epochs }}

        \rotatebox{90}{\put(27,-2.5){\color{black}{\footnotesize PSNR (dB) }}}
        \end{overpic}
    \end{minipage}

}
\hspace{4pt}
\subfigure[PSNR vs. Epochs ($\times2$)]
{
    \begin{minipage}[b]{.45\textwidth}
        \centering
        \begin{overpic}[scale=0.45]{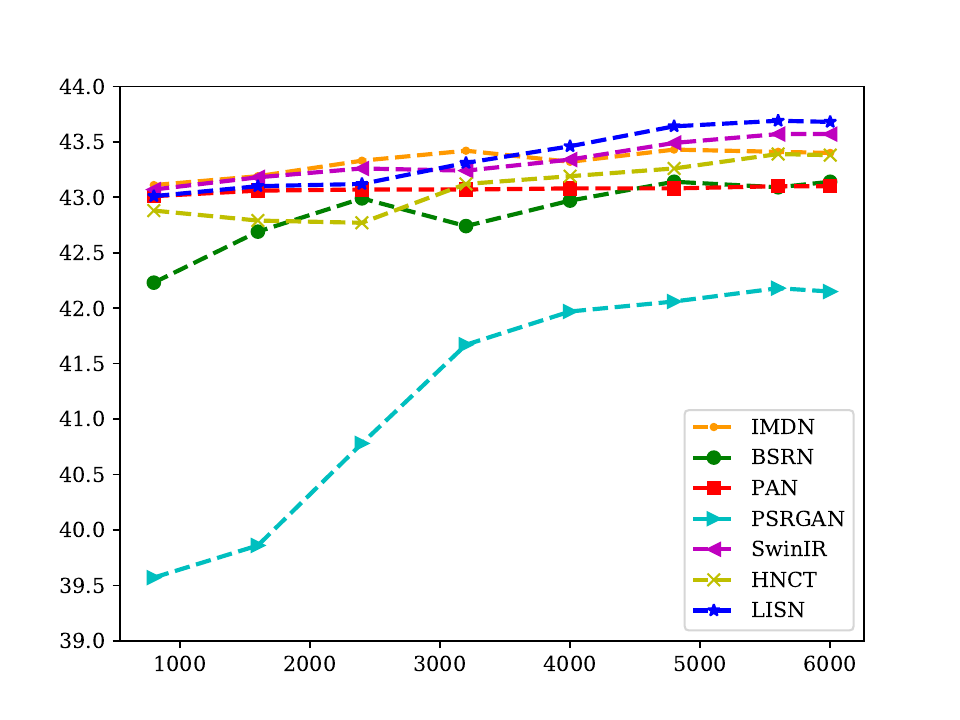}
        \put(25,-0.85){\color{black}{\footnotesize Number of training epochs }}
        \rotatebox{90}{\put(27,-2.5){\color{black}{\footnotesize PSNR (dB) }}}
        \end{overpic}
    \end{minipage}	
}
\caption{The effect of PSNR measure of the proposed model with respect to training epochs for scaling factors of 4 and 2.}\label{fig:psnr}
\end{figure*}

Furthermore, Fig.~\ref{fig:psnr} depicts the evolution of PSNR values over the training epochs for upscaling factors of $\times4$ and $\times2$. Fig.~\ref{fig: visual compare1} provides a visual comparison between LISN and other state-of-the-art methods on upscaling factors of $\times4$ and $\times2$. The test images were selected from the CVC-09-1K dataset. The magnified regions reveal that the infrared images reconstructed by LISN exhibit greater fidelity to the ground truth compared to competing methods, particularly in terms of edge and texture details. Remarkably, despite utilizing a relatively small number of parameters, LISN manages to achieve superior results in infrared image super-resolution. The visual comparisons further corroborate the effectiveness of our proposed LISN.

% PSNR图

% ablation table
\begin{table*}[t]
\begin{center}
\caption{Ablations of LISB for infrared image SR $(\times4)$ on the CVC-09-1K benchmark dataset. The default version represents the structure shown in Fig. 2.}\label{ablation-part}%
\begin{tabular}{@{}lccccc@{}}
\toprule
LISB & PSNR [dB] & Params [K] & FLOPs [G] &Time [ms] & Mem [M]\\
\midrule
No-split &\textbf{39.72} & 932 & 3.77 & 16.3 & 153.56  \\
No-RDB & 39.56& 278 & \textbf{1.06} & 16.2 & 90.64 \\
No-CCA & 39.59 & \textbf{277} & 1.08 & \textbf{15.9} & \textbf{87.57} \\
Default & 39.70& 279 & 1.10 & 16.4 & 93.64 \\
\bottomrule
\end{tabular}
\end{center}
\end{table*}

\subsection{Ablation Study}\label{subsec8}
To evaluate the efficacy of our proposed LISB, we performed a series of ablation studies on the CVC-09-1K benchmark dataset. The outcomes of these comprehensive experiments are detailed in Table \ref{ablation-part} and Table \ref{ablation-number}. In the subsequent section, we will explicate the influence of each discrete component within the framework.

\textbf{Channel Splitting Operation.} Our objective is to optimize the trade-off between performance and model complexity in the pursuit of efficient SR of infrared images. Toward this end, we have implemented a channel separation technique that is designed to segregate the feature map, thereby ensuring the maximal transfer of high-frequency details to the subsequent processing stage. As indicated in Table \ref{ablation-part}, the exclusion of the channel separation operation results in a negligible enhancement in the PSNR, increasing from 39.70 to 39.72. Nonetheless, the application of this methodology precipitated a significant, threefold escalation in the model's parameter count and Floating Point Operations Per Second (FLOPs), in stark contrast to the baseline model. 

% ablation number of LISB
\begin{table*}[htbp]
\begin{center}
\caption{Ablation study on the influence of the number of LISB.}\label{ablation-number}%
\scalebox{0.9}{
\begin{tabular}{@{}lccccc@{}}
\toprule
LISN & PSNR [dB] & Params [K] & FLOPs  [G] &Time [ms] & Mem [M]\\
\midrule
% \multicolumn{4}{l}{Role of LISB blocks (n)}\\
% n = 2  & 39.24   & \textbf{142} & \textbf{0.27}  \\
n = 4    & 39.53 & 232  & 0.93  & 16.2 & 87.53\\
n = 6 (Default)  & 39.70   & 279  & 1.10 & 16.4 & 93.64  \\
n = 8 & 39.72   & 397  & 1.58 & 31.1 & 98.38 \\
n = 10 & 39.73   & 480 & 1.90 & 38.2 & 103.56 \\
n = 12 & 39.75   & 563  & 2.23 & 44.8 & 108.23\\

\bottomrule
\end{tabular}
}
\end{center}
\end{table*}

\textbf{RDB and CCA.} The incorporation of the RDB within the LISB facilitates the refinement of local feature information by leveraging insights from the global feature map. In parallel, the CCA mechanism serves to bolster model efficacy by differentially weighting feature channels in accordance with their relevance to the specific task at hand. The ablation results are summarized in Table \ref{ablation-part}. As illustrated in the second row, the removal of the RDB component results in a decrement of 0.14 in the PSNR metric. Further, as demonstrated in the third row, the omission of CCA culminates in a 0.11 decline in the model's PSNR performance.

\textbf{Influence of the number of LISB.} To investigate the impact of the quantity of LISB on the ultimate super-resolution performance of the model, an ablation study was conducted, as shown in Table \ref{ablation-number}. The findings reveal a modest enhancement in model performance with an increase in the number of LISB. However, this improvement is accompanied by a significant escalation in the required parameters and FLOPs for the model. Consequently, in order to strike a balance between model complexity and performance, six LISBs were chosen for constructing the deep feature extraction module.

% 复杂度分析表
\begin{table*}[htbp]
\begin{center}
% \begin{minipage}{174pt} Model complexity comparison with other SR methods on a benchmark dataset (CVC-09-1K)
\caption{Model complexity comparison with other SR methods on results-C dataset. FLOPs is the abbreviation for floating point operations, which is evaluated on the LR image of size $64\times64$. The time represents the time it takes to process an image on an NVIDIA GeForce RTX 2080Ti GPU. "Mem" represents maximum GPU memory consumption during this test.}\label{complex-all}%
\small
\scalebox{0.9}{
% \begin{tabular}{\columnwidth}{@{\extracolsep{\fill}}lccccc@{\extracolsep{\fill}}}
\begin{tabular}{@{}lccccc@{}}
% \begin{tabular*}{columnwidth}{@{}lccccc@{}}
\toprule
Method & PSNR [dB] & Params [K] & FLOPs [G] &Time [ms]  & Mem [M]\\
\midrule
% EDSR    &  39.61  & 43,089  & 102.92  \\
IMDN \citep{hui2019lightweight}    & 33.82   & 715 & 3.46 & \textbf{11.3} & 102.46\\
BSRN \citep{li2022blueprint}  &  33.80  & 352  & 1.37 & 16.1 & 150.26 \\
PAN \citep{zhao2020efficient}  & 33.82   & \textbf{272}  & \textbf{1.01} & 14.7 & 133.31\\
PSRGAN \citep{huang2021infrared}  & 33.35   & 349  & 3.25 & 15.2 &250.48 \\
SwinIR \citep{liang2021swinir}  &33.85  & 897  & 6.94 & 142.9 & 252.31 \\
HNCT \citep{fang2022hybrid}  & 33.82   & 372  & 1.54 &34.5 & 156.71\\
\textbf{LISN(Ours) } & \textbf{33.86} & 279 & 1.10 & 16.4 & \textbf{93.64}\\
\bottomrule
\end{tabular}
}   
\end{center}
\end{table*}

\subsection{Model Complexity Analysis}\label{subsec9}
Given the constrained computing resources and limited memory capacity on infrared terminal devices, it is crucial to effectively manage the computational demands while optimizing the model's performance. As illustrated in Table \ref{complex-all}, our proposed LISN outperforms other state-of-the-art methods in attaining the highest super-resolution results for infrared images. Notably, the PSNR achieved by LISN exhibits a 0.51 improvement compared to PSRGAN \citep{huang2021infrared}. Moreover, LISN demonstrates minimal memory usage and notably lower FLOPs in comparison to other methods considered in the comparison.

\section{Conclusion}\label{sec13}
In this paper, we propose the LISN for enhancing the resolution of infrared images. The framework consists of four key components: shallow feature extraction, deep feature extraction, dense feature fusion, and HR infrared image reconstruction modules. To enable effective deep feature extraction, we employ a stack of LISBs. These LISBs combine a revamped Transformer and CNN, facilitating the extraction of a more suitable deep latent representation for super-resolution in infrared images. Moreover, the revamped transformer substitutes the conventional self-attention layer with a zero-parameter shift operation, significantly decreasing the computational burden of the LISB. Extensive experimental results demonstrate the superiority of our LISN over existing super-resolution methods, while also maintaining a remarkably low parameter count. In future work, we aim to explore the integration of this method with infrared small target detection to enhance the accuracy of detection.

\section*{Funding} This work was supported by "Pioneer" and "Leading Goose" R\&D Program of Zhejiang (No. 2023C03195), the Open Project Program of the State Key Laboratory of CAD\&CG (No. A2304), Zhejiang University, Aeronautical Science Foundation of China (No. 2022Z0710T5001), GuangDong Basic and Applied Basic Research Foundation (No. 2022A1515110570), Innovation Teams of Youth Innovation in Science, Technology of High Education Institutions of Shandong Province (No. 2021KJ088) and National Key Research and Development Program of China (No. 2023YFE0114900). The authors would like to thank the reviewers for their comments and suggestions in advance.

\backmatter

\bibliography{sn-article}% common bib file



\end{document}